# Magnetoresistance in Co-hBN-NiFe tunnel junctions enhanced by resonant tunneling through single defects in ultrathin hBN barriers

*Pablo U. Asshoff,[1,4] Jose L. Sambricio,[1,4] Sergey Slizovskiy,[1,4] Aidan P. Rooney,[2] Takashi Taniguchi,[3] Kenji Watanabe,[3] Sarah J. Haigh,[2] Vladimir Fal'ko,[1,4] Irina V. Grigorieva,[1,4]\* Ivan J. Vera-Marun [1,4]\**

[1]School of Physics and Astronomy, University of Manchester, Oxford Road, Manchester M13 9PL, UK

[2]School of Materials, University of Manchester, Oxford Road, Manchester M13 9PL, UK

[3]National Institute for Materials Science, 1-1 Namiki, Tsukuba 305-0044, Japan

[4]National Graphene Institute, University of Manchester, Manchester M13 9PL, UK

Hexagonal boron nitride (hBN) is a prototypical high-quality two-dimensional insulator and an ideal material to study tunneling phenomena, as it can be easily integrated in vertical van der Waals devices. For spintronic devices, its potential has been demonstrated both for efficient spin injection in lateral spin valves and as a barrier in magnetic tunnel junctions (MTJs). Here we reveal the effect of point defects inevitably present in mechanically exfoliated hBN on the tunnel magnetoresistance of Co-hBN-NiFe MTJs. We observe a clear enhancement of both the conductance and magnetoresistance of the junction at well-defined bias voltages, indicating resonant tunneling through magnetic (spin-polarized) defect states. The spin polarization of the defect states is attributed to exchange coupling of a paramagnetic impurity in the few-atomic-layer thick hBN to the ferromagnetic electrodes. This is confirmed by excellent agreement with theoretical modelling. Our findings should be taken into account in analyzing tunneling processes in hBN-based magnetic devices. More generally, our study shows the potential of using atomically thin hBN barriers with defects to engineer the magnetoresistance of MTJs and to achieve spin filtering, opening the door towards exploiting the spin degree of freedom in current studies of point defects as quantum emitters.

Utilizing two-dimensional (2D) materials to create functional devices offers many exciting opportunities for electronics [1]. The assembly process of these tailored van-der-Waals heterostructures is relatively easy and does not involve standard problems of thin film deposition, such as island growth and low crystallinity. Instead, individual atomically thin layers of high crystalline quality can be integrated into a device by a simple transfer process, allowing the physical properties of the device to be accurately designed [1,2,3,4]. In the last decade 2D materials have also had a significant impact on spintronics, a subfield of electronics where the spin degree of freedom is exploited, including demonstrations of new functionalities in magnetic tunnel junctions (MTJs) [2,3,5,6,7,8,9,10]. MTJs, widely used as sensors, are ferromagnet/insulator/ferromagnet (FM/I/FM) heterostructures where the nature of FM electrode/barrier interfaces is known to critically affect the central metric of device performance, its magnetoresistance (MR), in myriad ways [11]. For instance, if the barrier is crystalline and lattice-matched with the ferromagnets, spin filtering may occur (i.e., electrons with a particular spin state are transported preferentially) [2,11,12,13]. Alternatively, bonding or hybridization of the atomic orbitals at the FM/barrier interfaces can be the dominant factor that determines spin-dependent tunneling [2,11] while defects, impurities and pinholes in the barrier can open additional conduction channels, modifying the MR [14,15,16,17,18,19,20,21,22]. Using 2D materials as an insulating barrier between the FM electrodes is of particular interest, with atomically thin hexagonal boron nitride (hBN) currently being the material of choice: mono- or bilayer hBN has been shown to provide efficient spin-injection in graphene-based lateral spin-valves [3,23] and a few percent MR has been observed in MTJs with atomically thin hBN barriers [2,5,6].

In this report we focus on the role and potential exploitation in MTJs of point defects that are inevitably present in atomically thin hBN. The presence of such defects in the best-quality, mechanically exfoliated hBN has been demonstrated in several recent studies, e.g. nitrogen and boron vacancies [24,25] and carbon and oxygen impurities [26,27], some of which are believed to be paramagnetic [26,24]. Spatially separated and oppositely charged single-site defects have been imaged by scanning tunneling microscopy [26]. In another report [28], energy-resolved spectroscopic features of defect-mediated tunneling were seen in transport measurements on non-magnetic Cr/Au-hBN-Cr/Au tunnel junctions with a ~6-layer-hBN barrier. In the case of MTJs, point defects in hBN, especially if they are magnetic (spin-polarized), can be expected to assist or otherwise affect spin-dependent tunneling and, therefore, the MR of the device. In particular, theory predicts that resonant tunneling through magnetic impurities should lead to a notable increase in MR [14,29,30], whereas non-magnetic impurities can be expected to reduce it [14,15,31]. This has been a topic of high interest in the literature, but so far mostly limited to theoretical studies with scarce direct experimental evidence. While there has been significant experimental effort to understand the effect of defect-mediated transport in conventional MTJs (e.g., due to Si, Ni [16], Fe [17], Gd or Dy [18] dopants introduced into the $AlO_x$ barrier), the results regarding their impact on the MR remained largely inconclusive, due to averaging over large numbers of defects. In contrast, atomically thin hBN creates unrivalled opportunities for the experimental study of the effect of well-separated - both spatially and in terms of energy - point defects on spin-polarized transport. Conversely, as point defects in hBN have been shown to act as quantum emitters of single photons [32], controlling their spin degree of freedom opens a pathway for quantum spintronics [33,34].

A schematic view of the MTJs used in our study is shown in Fig. 1a: the device consists of two FM electrodes (Co and $Ni_{0.8}Fe_{0.2}$) separated by 2-4 atomic layer-thick hBN mechanically exfoliated from bulk hBN crystals. We used hBN from the same supplier as in refs. [26,28] and therefore can expect to see a finite density of isolated point defects. To ensure clean interfaces between the FM electrodes and hBN crystals, we used a fabrication technique developed in our previous study [2] where the ferromagnetic metals were deposited on the two sides of a suspended hBN membrane, thereby preventing oxidation of the ferromagnets, minimizing the number of fabrication steps and limiting the exposure of the interfaces to solvents during preparation. We note that oxidation and degradation of the interfaces is a fundamental problem if a conventional bottom-up MTJ fabrication approach is used, which involves transfer of the 2D material onto a ferromagnetic film [6]. Details of our fabrication procedure can be found in ref. [2]. Briefly, an exfoliated few-layer hBN flake was transferred onto a 3-4 μm diameter circular aperture in a 100 nm-thick $SiN_x$ membrane using a dry transfer method [35]. The number of layers was estimated from optical contrast, using differential interference contrast microscopy. After that, 20 nm-thick Co and $Ni_{0.8}Fe_{0.2}$ films were evaporated in vacuum ($10^{-6}$ mbar base pressure) onto the suspended flake from the top and bottom side, respectively. While the sample was briefly exposed to air between the two steps of metal deposition on the top and bottom of the hBN, oxidation of the ferromagnets was avoided as they were protected by hBN on one side and a Ti/Au capping on the other. To ensure good contact between the hBN spacer and ferromagnetic electrodes, the devices were annealed for several hours at 300°C in $Ar/H_2$ atmosphere, a standard procedure known to eliminate residual contamination and result in clean, uniform interfaces between the 2D crystal and the FM films [2].

To reveal the contribution of the defects in hBN to the spin-dependent transport, we used bias spectroscopy, i.e., studied magnetic-field dependent vertical transport either under a pure *dc* bias, $V_b$, or a small *ac* excitation superimposed on the *dc* bias, with detection of the *ac* components of the junction voltage d$V$ and current d$I$ by a lock-in amplifier. Shown below are the results obtained on two typical devices (similar data for two more devices are shown in Supporting Information Section 4). Most of the results correspond to a MTJ with a 3-layer hBN barrier, where we found clear signatures of defect-mediated resonant tunneling. For comparison, we also show a 4-layer hBN device where no contribution from defect states could be seen. A typical *I-V* characteristic is shown in the inset of Fig. 1b: As expected for a MTJ, it is non-linear, with the non-linearity visible more clearly in the $R(V_b)$ and d$V$/d$I(V_b)$ plots in Fig. 1c. Comparison of the resistance × area (RA) product of this device (1.53 MΩ·μm$^2$ measured for a junction area of 7.3 μm$^2$) with the known dependence of the RA product on the thickness of the hBN tunnel barrier [4] allowed us to determine the hBN barrier thickness, 3 atomic layers in this case. The inset of Fig. 1c shows a magneto-transport characteristic of the device, consistent with classical MTJ behavior: It is clear that the 3-layer hBN is sufficiently thick to decouple the two FM layers that switch independently, without any evidence of magnetic exchange coupling between the electrodes, and with a maximum and minimum resistance corresponding to the antiparallel (AP) and parallel (P) configuration, respectively. In the upper panel of Fig. 1c, the resistance $R$ is shown as a function of bias voltage $V_b$ for the AP and P configuration. The resulting magnetoresistance MR = $(R_{AP} - R_P)/ R_P$ is shown in Fig. 2a (black curve). For comparison, Fig. 2a also shows corresponding data for the device with a 4-layer hBN barrier. The maximum MR is +3.6 % and +2.2 % for the 3-layer and 4-layer hBN device, respectively. This compares favorably to previously reported MR ~ 0.4% [6] and ~6% [5], where CVD-grown monolayer hBN was used and, as noted by the authors, the fabrication process involved exposure of the FM layers to air, likely resulting in FM oxidation and effectively adding an extra insulating layer. The monotonic decrease of MR for both devices in Fig. 2a with increasing $V_b$ is a standard behavior observed in MTJs due to spin-flip scattering [11,36], whereas the asymmetry in the bias dependence can be expected for dissimilar electrode materials, as in our case [6,37].

To further elucidate the behavior of our MTJs, we measured the temperature dependence of the junction resistance, $R(T)$, which was not reported for tunnel junctions with an hBN barrier in earlier studies. Surprisingly, this showed that $R(T)$ is metallic-like, decreasing by ~20% from room temperature to ~50 K and becoming almost temperature independent between 50 and 10 K – see Fig. 1b. This is in contrast to expectations for thermally assisted transport through an insulating barrier but is similar to $R(T)$ behavior observed previously for $AlO_x$ and MgO-based MTJs, where it was attributed to defect-mediated conduction through the barrier [19,20]. This $R(T)$ behavior is presently observed for all hBN-based devices (see Supplementary Information Section 2).

To obtain additional information about possible conduction through defect channels, we measured the differential resistance $dV/dI$ shown in the lower panel of Fig. 1c, and calculated the corresponding differential MR (dMR) defined as dMR = $(dV/dI_{AP} - dV/dI_P)/ dV/dI_P$ - see Fig. 2b. While the $R(V_b)$ curve (top panel of Fig. 1c) reflects the total current through the junction, $dV/dI(V_b)$ and dMR are sensitive to the local electronic configuration at the Fermi level, $E_F$, and to spin-dependent changes in a narrow energy range around $E_F$, respectively [38]. Therefore, any defect-related resonant tunneling is expected to show up more clearly in the $dV/dI(V_b)$ and dMR$(V_b)$ traces. Indeed, $dV/dI(V_b)$ in Fig. 1c exhibits several features, most prominently a feature at $V_b \approx +0.12$ V. Most of these features do not appreciably affect the MR, indicating their non-magnetic origin. This can be seen in Fig. 2a for both devices presented here (3-layer and 4-layer hBN barriers). However, the dip in the tunneling resistance $R(V_b)$ of the device with 3-layer hBN at $V_b$ =+0.12 V (enhanced differential conductance) is associated with a small but clear peak in the magnetoresistance. The latter is visible in Fig. 2a and becomes particularly prominent in the differential MR (Fig. 2b). Relative to the background magnetoresistance in the relevant range of $V_b$ (between +0.1 and +0.14 V) the MR increases by ~3% (or 0.1% in absolute terms, from MR=~3.0% to MR=~3.1% at the left side of the feature), whereas the dMR increases by up to ~50% (from dMR=~2% to dMR~3%).

To shed light on the origin of different features on d$V$/d$I$ (or d$I$/d$V$), we analyze the $I$-$V$ characteristic in the full bias range by studying its second derivative, $d^2I/dV^2$. The resulting inelastic electron tunneling spectra (IETS), shown in Fig. 3a, exhibit a prominent antisymmetric response with a series of peaks at similar positions in negative and positive bias. These features are consistent with van Hove-like peaks in the single phonon density of states of hBN at energies between ~10 and 190 meV, as found both theoretically and experimentally [39,40]. Accordingly, we attribute the $d^2I/dV^2$ peaks in Fig. 3a to inelastic tunneling enabled by phonon emission in the hBN barrier. Next, we extracted the symmetric component of the spectra with respect to the magnetization direction of the FMs (parallel–antiparallel) – this is shown in the bottom panel of Fig. 3b. This showed that there is only one magnetic-field-dependent feature, localized at $V_b$ ≈+0.12 V, which consists of both peaks and valleys. Such behavior is characteristic of elastic trap-assisted tunneling via defects [28] and is in contrast to inelastic tunneling processes that appear only as peaks [41]. The magnetic-field dependence of this feature is fully consistent with our observation that enhancement of the magnetoresistance only occurs at $V_b$ ≈+0.12 V.

By subtracting the symmetric component from the raw IETS, we obtained the spectrum shown in the top panel of Fig. 3b. This antisymmetric response is similar for both magnetization states and consists of only peaks. For a quantitative comparison with hBN-based vertical tunneling devices from literature, we performed a multi-peak Gaussian fit of the form $\exp(-E^2/2\gamma^2)$ and compared the results with the known positions of inelastic phonon peaks for graphene-hBN-graphene heterostructure with dominant hBN features (Device 2 in ref. [39]). Our multi-peak analysis yielded an excellent fit to the data, with an energy broadening $\gamma = 8$ meV for all peaks. Most importantly, all fitted peaks are located within ~10 mV of the bias values reported in ref. [39], with an excellent agreement within ~3 mV for the four main peaks labelled (v) to (viii) and located at $V_b < 0.10$ V (see Supplementary Information Section 3). We note that peak (viii) at ~15 mV corresponds to low-energy phonons close to the Γ point of hBN and has been the object of studies by inelastic x-ray spectroscopy [40], whereas peak (vii) at ~37 mV is associated with the lowest-energy acoustic mode in the vicinity of the M and K points [39]. The rest of the peaks are all related to prominent features in the phonon density of states, with the high-bias peaks (i) and (ii) located at $V_b > 0.15$ V associated with high-energy optical phonons of hBN [40]. A similar IETS was obtained for the reference 4-layer device (see Supplementary Information Section 3). The above analysis allows us to conclude that, except for the magnetization-dependent feature at +0.12 V, all prominent low-energy peaks are associated with phonon-assisted tunneling that do not enhance the MR. This also provides further evidence that our fabrication process ensured high-quality FM/hBN interfaces and that transport is dominated by tunneling through hBN rather than, e.g., a contamination layer.

To address the observed enhancement of both the conductance (Fig. 1c) and MR (Fig. 2) around +0.12 V, we recall that sharp features in conductance characteristics (*I-V*, d*I*/d*V* and $d^2I/dV^2$) of MTJs with traditional oxide barriers are known to be signatures of trap-assisted tunneling due to in-gap defect states in e.g. MgO or $HfO_2/Y_2O_3$ barriers [21,41]. We show in the bottom panel of Fig. 4a that our observations of the enhanced conductance (peak in $dI/dV$ at $V_b$=+0.12 V), and the corresponding features in $d^2I/dV^2$ (bottom panel of Fig. 3b), closely match those reported for defect-assisted tunneling in refs. [21,41]. The fact that the dip in $dV/dI$ and the corresponding increase in the MR of our device are observed in the narrow range of $V_b$ and only at low $T$ is also consistent with defect-assisted tunneling, as the latter requires alignment of the Fermi energy with the energy of the localized defect state. Indeed, the sharp oscillation in the differential MR at 10 K is strongly suppressed at 50 K (top panel of Fig. 4a and inset), as expected for defect-mediated transport [29].

Regarding the origin of these localized defect states, we can exclude degradation of the junction or formation of pinholes due to dielectric breakdown, as $V_b$ was kept well below the known maximum limit ~ 0.7 V/nm for hBN [42]. Furthermore, as the fabrication procedure involved only solvent-free, non-degrading processes, we attribute these localized states to pre-existing barrier properties, not affected by the fabrication process. It is therefore natural to attribute the conductance and MR features to the defects/impurities always present in hBN, as discussed above and in refs. [26,28].

Having established that the feature at +0.12 V must be related to defect-mediated tunneling, the increased MR suggests that the defect state in the barrier is magnetic (spin-polarized) [14,29]. An intuitive picture of the underlying physical process, depicted in Fig 4b, has been described in ref. [30]: The increased MR can be understood by decomposing transport through a defect-containing MTJ into two processes, firstly from the first FM electrode into the magnetic defect, and secondly from the defect into the second electrode, effectively acting as two MTJs in series and enhancing the MR. The small thickness of the hBN barrier (3 atomic layers) is likely to play a crucial role in observing the increased MR in our MTJ, as an enhanced magnetic coupling between the defect and each of the ferromagnetic electrodes will result in spin-split defect states, even if the impurity is *a priori* non-magnetic (e.g., paramagnetic oxygen atom or a boron monovacancy [25]). Such magnetic coupling between a paramagnetic impurity and the neighboring FM electrode was analyzed theoretically in ref. [29], predicting spin polarization of the tunneling electrons at resonance with the impurity state and an enhanced MR.

To further understand the nature of the transport process in our MTJ through the defect appearing at $V_b \approx +0.12$ V, we compare our data with the known theoretical model where the contribution of resonant tunneling through a defect to the total conductance across the barrier is given by [14,43]

$$G(\epsilon) = \frac{4e^2}{h} \sum_{\mu=\uparrow,\downarrow} \frac{\Gamma_\mu^L \Gamma_\mu^R}{\left(\epsilon - \epsilon_{i,\mu}\right)^2 + \left(\Gamma_\mu^L + \Gamma_\mu^R\right)^2}$$

Here $\epsilon$ is the energy of the tunneling electron, $\epsilon_{i,\mu}$ the energy of the quasi-bound state of the electron near the defect, and $\Gamma^{L/R}$ are partial widths of the resonance corresponding to electron tunneling between the defect and the left/right FM electrode. The model assumes that the energy levels at the defect are spin-split – see Supporting Information for a detailed description. We find an excellent agreement between the calculated dI/dV and the experimental data, as shown by the fit in the bottom panel of Fig. 4a. The splitting of the defect levels is larger for parallel magnetization of electrodes, when the coupling to spin polarization of both electrodes adds up, increasing the separation between the corresponding spectroscopic peaks. This clearly indicates that the defect is magnetic. As suggested in ref. [29], the splitting of the defect energy levels may be caused by exchange or super-exchange interaction of the paramagnetic impurity with spin-polarized electrons in the FM electrodes, becoming significant for defect levels in narrow insulating barriers, as in our case. A similar scenario has been reported in ref. [44] where tunneling spectroscopy experiments suggested that spin-split impurity states in a 2.5 nm MgO barrier were exchange coupled to the FM electrodes. In our case with an ultrathin 3-layer hBN barrier (thickness $\approx$1 nm), the atomic thickness facilitates exchange coupling of a single defect to both electrodes, and therefore the energy of the defect state can be tuned by the relative magnetic alignment of the electrodes, in contrast to thicker barriers − see Supporting Information for a detailed discussion.

The different spin splitting and bias dependences for parallel and antiparallel magnetization, as shown in the bottom inset of Fig. 4a, should lead to a sharp peak in MR at +0.12 V followed by a dip at higher $V_b$ (where the conductance peak for the anti-parallel magnetization is located), as indeed observed. At elevated temperatures the inelastic processes and thermal broadening start to dominate, which explains that the defect-mediated peak disappears in the experiment at 50 K.

Using the fitting parameters for the theory curves in Fig. 4a we were able to extract information about the location of the defect and the proximity-induced spin splitting, $\Delta^P \approx 9.8$ meV, $\Delta^{AP} \approx 7.0$ meV, in good agreement with theory [45] (the latter depends on the relative magnetization of the FMs, see Supporting Information). As the fit yielded $\Gamma^L \ll \Gamma^R$, the defect must be located asymmetrically, probably between two of the three hBN layers, rather than at the center of the barrier. This explains the observed asymmetry in the MR response for positive and negative $V_b$. Furthermore, comparing the contributions to spin splitting from the two FM electrodes showed that it must decay with distance at a slower rate (~ 1/5 per hBN layer) compared to the tunneling conductance (~ 1/50 per hBN layer, see Supporting Information). Such a slow decay of proximity-induced spin splitting at hBN/(Co,Ni) interfaces (~ 1/10 per hBN layer) has been predicted theoretically in a recent work [45] and, to the best of our knowledge, our observations are the first experimental evidence of this effect.

We emphasize that the discussed conductance/MR feature at $V_b \approx +0.12$ V is well separated from inelastic phonon-assisted peaks (see above) and most likely corresponds to tunneling through a *single* defect. [28]. Furthermore, the observed MR enhancement by ~3% seems modest but compares very favorably with the maximum enhancement predicted by theory [14]. In the latter case, the maximum difference between direct tunneling and resonant tunneling was shown to depend on the position of the defect inside the barrier, and for the simplest case with a defect located in the center of the barrier was found to be ~13% for relatively thick, > 2 nm, barriers, where direct tunneling is significantly suppressed and conductance is dominated by resonant tunneling. In our ~1 nm thick MTJ direct tunneling remains a major contribution, which explains the relatively modest enhancement of both the junction conductance and the MR. We note that it is not surprising that we observe only one resonant feature, even though one can expect a number of defects to be present in the hBN barrier. This is because a significant contribution of impurity-assisted resonant tunneling can only be expected for optimum conditions: defects close to the center of the barrier and to the Fermi level [46]. Another promising route to amplify the relative contribution of the defects is by employing nanojunctions with a very small area: Here the effect on the MR is expected to be much larger and a single impurity may lead to the MR sign inversion [22,47], as was seen experimentally in Ni/NiO/Co junctions with < 0.01 $\mu m^2$ area [36].

In summary, we report experimental observations of enhanced conductance and magnetoresistance in a high-quality, crystalline hBN barrier in Co-hBN-NiFe MTJs, at well-defined bias voltages. As demonstrated by comparison with theory, our results correspond to resonant tunneling through a magnetic (i.e., spin-polarized) defect. Our findings demonstrate the potential of using few-layer hBN crystals as an outstanding platform for control of the tunnel magnetoresistance through introduction or isolation of single defects. For example, it should be possible to enhance the MR increase using thicker hBN crystals where direct tunneling is strongly suppressed but the defect state(s) remain magnetically aligned with the FM electrodes. Furthermore, as suggested in ref. [14], hBN-based MTJs may enable the observation of MR with only one magnetic electrode, as the magnetic defects in the barrier will act as spin filters. Control of spin transport via individual defect states would enable a 2D-based platform for circularly-polarized single-photon emitters [32,33] for quantum spintronics [34].

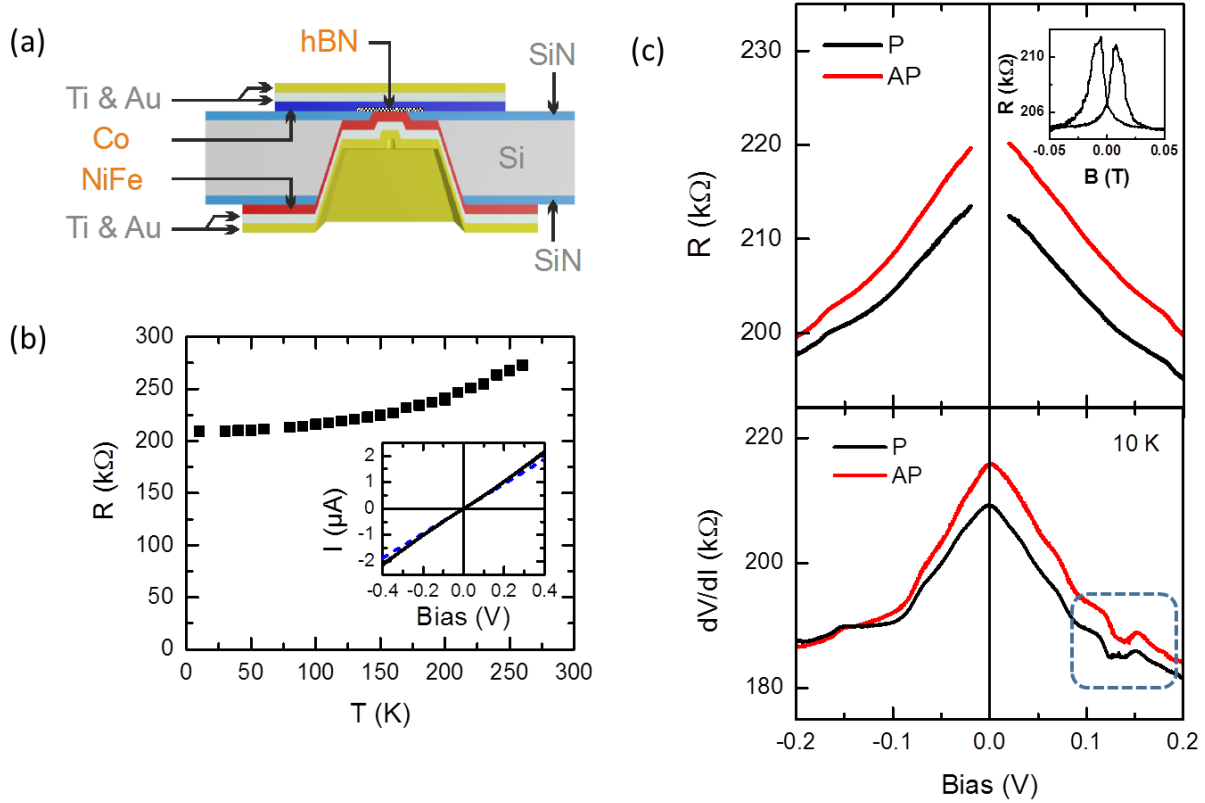

FIG. 1. Device schematic and the resistance of the 3-layer hBN sample. (a) Design of the NiFe-graphene-Co samples. First, hBN is suspended over an aperture in SiN, followed by the evaporation of the ferromagnetic electrodes Co and NiFe from the top and bottom, respectively. Ti, Au are capping layers to prevent oxidation of the ferromagnets. (b) $R(T)$ dependence of the junction resistance. ($V_b$ = 20 mV, parallel configuration). Inset: I-V characteristic of the sample ($T$ = 10 K). The tangent to the curve at zero bias, shown as a dashed line, clearly diverges from the data for higher voltages. (c) Bias spectroscopy ($T$ = 10 K). Top panel: $R(V)$ for the parallel (P) and antiparallel (AP) configuration of the ferromagnetic electrodes. Inset: Typical MR trace, $V_b$ = +90 mV. Bottom panel: Differential resistance d$V$/d$I$($V_b$) of the MTJ, with a feature at +0.12 V circled (P/AP: parallel/antiparallel alignment of the ferromagnetic layers).

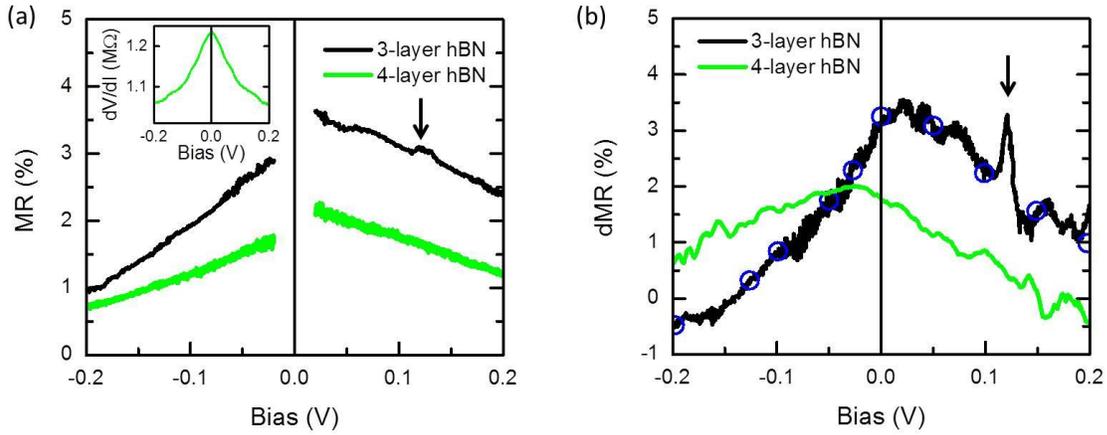

FIG. 2. **Magnetoresistance of the junctions as a function of bias voltage**. (a) DC magnetoresistance MR = $(R_{AP}-R_P)/R_P$ of the 3-layer hBN (black) and 4-layer hBN (green) device with maximum MR of +3.6 % and +2.2 %, respectively. Data for the 3-layer hBN device has been derived from $R_P$, $R_{AP}$ in the top panel of Fig. 1c. Inset: Differential resistance (d$V$/d$I$) of the 4-layer device. (b) Differential MR (see text) for the same devices as in (a). Data for the 3-layer hBN device (black) has been derived from d$V$/d$I_P$, d$V$/d$I_{AP}$ in the bottom panel of Fig. 1c. Data points shown as open symbols are obtained from individual traces of differential MR vs magnetic field at fixed $V_b$. A strong change of (differential) MR occurs at ~+0.12 V for the 3-layer device (marked by the vertical arrow). All data taken at $T = 10$ K.

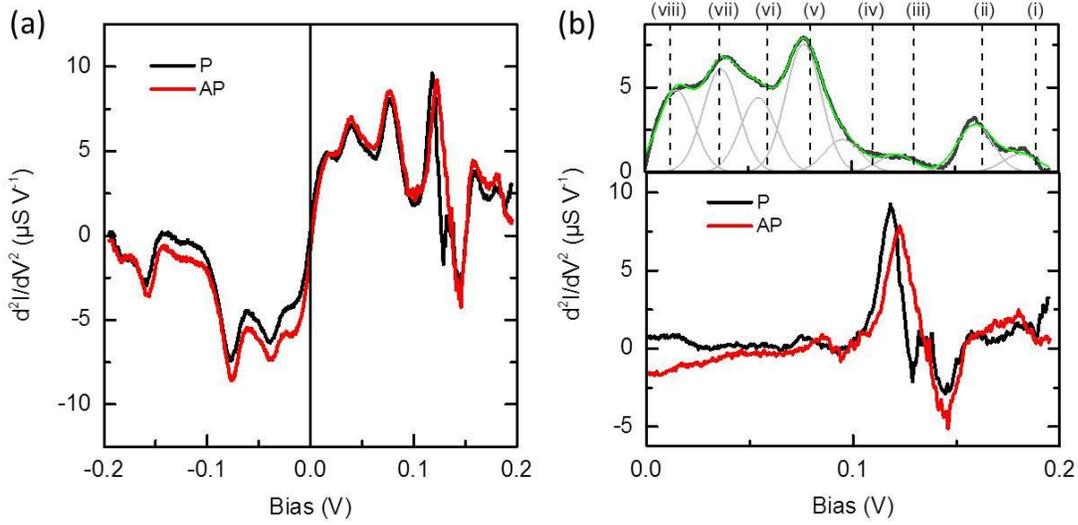

FIG. 3. **Full bias range analysis via inelastic electron tunneling spectroscopy (IETS).** (a) Raw $d^2I/dV^2$, numerical derivative of $dI/dV$ (P/AP: parallel/antiparallel alignment of the ferromagnetic electrodes). (b) Bottom panel: symmetric IETS component for both magnetic configurations, obtained by adding the contributions for the two bias polarities. Top panel: After removal of the symmetric component of raw IETS. The response for both magnetic configurations is very similar, so they have been averaged for clarity (black line). A multi-peak fit (green line) and its individual peaks (gray lines) are shown. Vertical dotted lines, labelled (i) to (viii), indicate the position of phonon peaks reported in ref. [39].

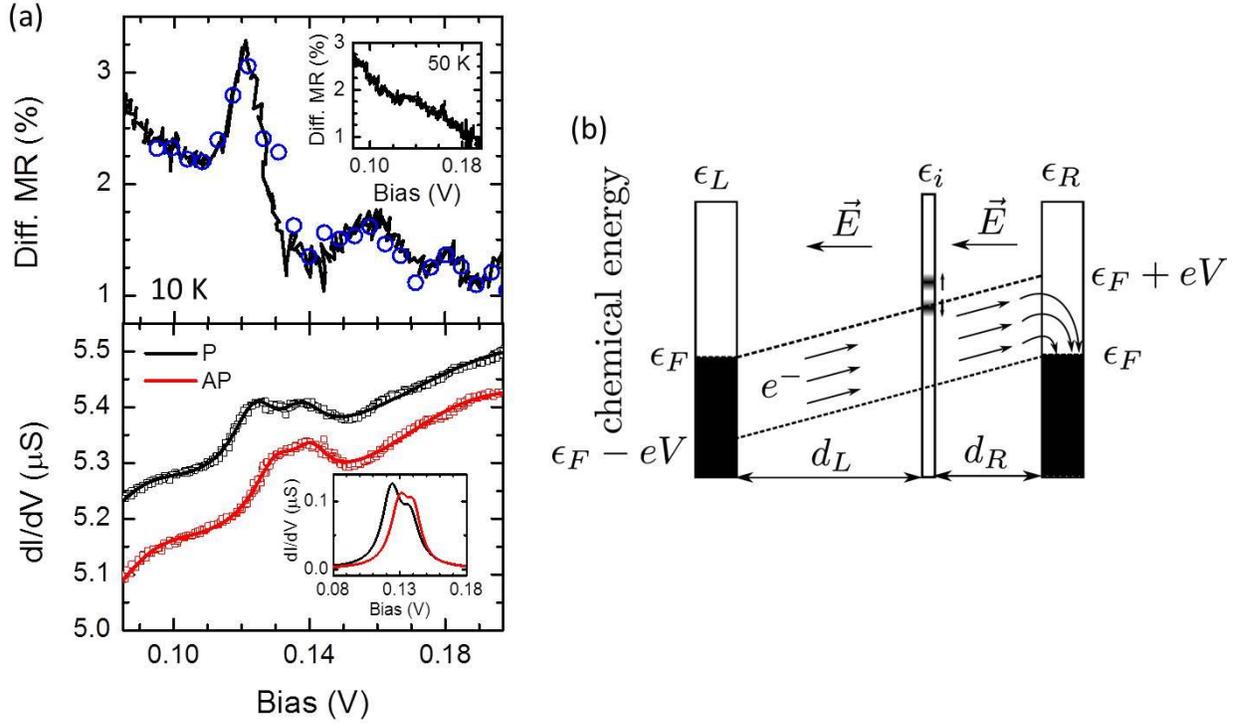

FIG. 4. **Enhanced magnetoresistance around +0.12 V and its comparison with theory. (a) Top panel:** Differential MR (see text) at 10 K. Open symbols are data points obtained from individual traces of differential MR vs magnetic field at fixed $V_b$. **Top inset:** Differential MR at 50 K. **Bottom panel:** conductance (d$I$/d$V$) at 10 K. Open symbols are experimental data (zoom of the circled part of the d$V$/d$I$ curves in the bottom panel of Fig. 1b presented here as differential conductance). Solid lines are fit to the theoretical model. **Bottom inset:** Calculated contributions to the conductance from the magnetic defect. The splitting of the P and AP curves can be clearly seen. **(b)** Schematic of the theoretical model describing electron tunneling through an insulating barrier containing a defect under an applied bias voltage $V$. Electrons on the left electrode are accelerated by the electric field $\vec{E}$ and tunnel following a path through the defect location. $d_{L/R}$ are distances from the defect to the left/right electrode, respectively.

## ASSOCIATED CONTENT

**Supporting Information**. The Supporting Information contains details on the theoretical model and quantitative analysis of the data, the temperature dependence of junction resistance for all samples, further details of the inelastic electron tunneling spectroscopy, and data for two more devices, one reference and one with resonant features (PDF).

## AUTHOR INFORMATION

**Corresponding Authors**

*E-mail: irina.grigorieva@manchester.ac.uk

*E-mail: ivan.veramarun@manchester.ac.uk

## ACKNOWLEDGMENT

We acknowledge support from the EC-FET Graphene Flagship, grant agreement no. 604391, FP7 FET-Open grant 618083 (CNTQC) and from the FP7 Marie Curie Initial Training Network 'Spintronics in Graphene' (SPINOGRAPH), grant 607904.

[3] Kamalakar, M. V., Dankert, A., Bergsten, J., Ive, T. & Dash, S. P. Enhanced tunnel spin injection into graphene using chemical vapor deposited hexagonal boron nitride. *Sci. Rep.* **2014**, *4*, 6146.

[4] Britnell, L. *et al*. Electron tunneling through ultrathin boron nitride crystalline barriers. *Nano Lett.* **2012**, *12*, 1707-1710.

[5] Piquemal-Banci, M. *et al*. Magnetic tunnel junctions with monolayer hexagonal boron nitride tunnel barriers. *Appl. Phys. Lett.* **2016**, *108*, 102404.

[6] Dankert, A., Kamalakar, M. V., Wajid, A., Patel, R. S. & Dash, S. P. Tunnel magnetoresistance with atomically thin two-dimensional hexagonal boron nitride barriers. *Nano Res.* **2015**, *8*, 1357-1364.

[7] Roche, S. *et al*. Graphene spintronics: the European Flagship perspective. *2D Mater.* **2015**, *2*, 030202.

[8] Wang, Z. *et al*. Very large tunneling magnetoresistance in layered magnetic semiconductor CrI3. *Nature Communications* **2018**, *9*, 2516.

[9] Huang, B. *et al*. Electrical control of 2D magnetism in Bilayer CrI3. *Nature Nanotechnology* **2018**, *13*, 544–548.

[10] Ghazaryan, D. *et al.* Magnon-assisted tunnelling in van der Waals heterostructures based on CrBr3. *Nature Electronics* **2018**, *1*, 344-349.

[11] Tsymbal, E. Y. *et al.* Spin-dependent tunnelling in magnetic tunnel junctions. *J. Phys.: Condens. Matter* **2003**, *15*, R109-R142.

# SUPPORTING INFORMATION

# Magnetoresistance in Co-hBN-NiFe tunnel junctions enhanced by resonant tunneling through single defects in ultrathin hBN barriers

P. U. Asshoff,[1,4] J. L. Sambricio,[1,4] S. Slizovskiy,[1,4] A.P. Rooney,[2] T. Taniguchi,[3] K. Watanabe,[3] S. J. Haigh,[2] V. Fal'ko,[1,4] I. V. Grigorieva,[1,4*] I. J. Vera-Marun [1,4*]

[1]*School of Physics and Astronomy, University of Manchester, Oxford Road, Manchester M13 9PL, UK*

[2]*School of Materials, University of Manchester, Oxford Road, Manchester M13 9PL, UK*

[3]*National Institute for Materials Science, 1-1 Namiki, Tsukuba 305-0044, Japan*

[4]*National Graphene Institute, University of Manchester, Manchester M13 9PL, UK*

**Corresponding Authors**

*E-mail: irina.grigorieva@manchester.ac.uk

*E-mail: ivan.veramarun@manchester.ac.uk

## 1. Modelling of resonant tunneling through single spin-split defect in ultrathin hBN barriers

We consider resonant tunneling of electrons from the left electrode through defect states in the hBN barrier to the right electrode. The contribution of this process to the total conductance across the barrier is given by [S1]

$$G(\epsilon) = \frac{4e^2}{h} \sum_{\mu=\uparrow,\downarrow} \frac{\Gamma_\mu^L \Gamma_\mu^R}{(\epsilon - \epsilon_{i,\mu})^2 + (\Gamma_\mu^L + \Gamma_\mu^R)^2} \tag{1}$$

where $\epsilon$ is the energy of the electron, $\epsilon_{i,\mu}$ is the energy of the quasi-bound state of the electron near the defect $i$ for spin state $\mu$, and $\Gamma^{L/R}$ are the partial widths of the resonance, corresponding to electron tunneling between the defect and the left/right electrode. We choose to work in terms of chemical energy, considering the electric field contribution separately. In this approach the chemical energy of the defect level is not affected by the bias voltage, see Fig. S1. With an applied bias voltage $V$ and assuming elastic tunneling, we have electrons in the range $(\epsilon_F - eV, \epsilon_F)$ on the left electrode, which are accelerated by the electric field to energies $(\epsilon_F, \epsilon_F + eV)$ at the right electrode. These electrons are passing through the plane where the defect is located with energies in the range $(\epsilon_F + eV\frac{d_L}{d_L + d_R} - eV, \epsilon_F + eV\frac{d_L}{d_L + d_R})$, where $d_{L/R}$ are distances from the defect to the left/right electrode.

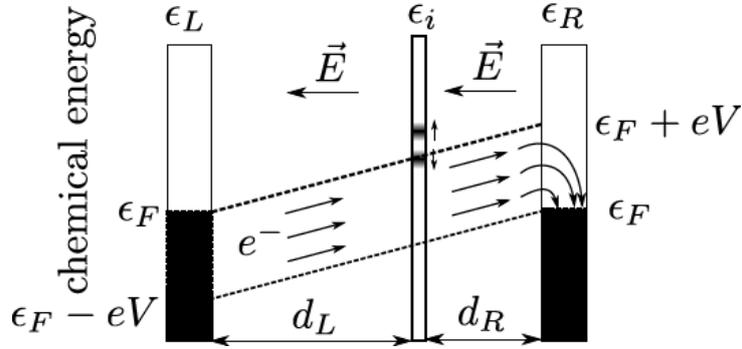

FIG. S1. Tunneling between the defect and the electrodes under an applied bias voltage $V$.

The resulting contribution of one defect to the current is

$$I(V) = e^{-1} \int_{\epsilon_F - eV}^{\epsilon_F} d\epsilon\, G\left(\epsilon + eV\frac{d_L}{d_L + d_R}\right) = e^{-1} \int_{\epsilon_F - eV\frac{d_R}{d_L + d_R}}^{\epsilon_F + eV\frac{d_L}{d_L + d_R}} d\epsilon\, G(\epsilon). \tag{2}$$

Differentiating w.r.t. $V$ and assuming that the resonance is located near the upper limit of integration leads to

$$\frac{dI}{dV} = \frac{4e^2 d_L}{h(d_L + d_R)} \sum_{\mu=\uparrow,\downarrow} \frac{\Gamma_\mu^L \Gamma_\mu^R}{\frac{e^2 d_L^2 (V - V_{i,\mu})^2}{(d_L + d_R)^2} + (\Gamma_\mu^L + \Gamma_\mu^R)^2}, \tag{3}$$

where $V$ lies near $V_{i,\mu}$, defined as $\epsilon_{i,\mu} = \epsilon_F + eV_{i,\mu}\frac{d_L}{d_L+d_R}$. We note that, as illustrated in Fig. S1, a defect contributes to the conductance only if its energy lies within the interval corresponding to the limits of integration in (2), i.e. its contribution is bias-dependent and, unless its energy is close to $\epsilon_F$, requires a sufficiently large bias to have an effect on conductance.

Having defined the defect's contribution in this way, we can fit the data for differential conductance in Fig. 4a with a combination of a smooth curve (corresponding to direct tunneling/less-dominant phonon-assisted tunneling, as discussed in the main text) and Lorentzian peaks (defect-assisted resonant tunneling):

$$\frac{dI}{dV} = \frac{a_1 V + a_2}{1 + a_3 V} + \sum_i \frac{c_i \gamma_i^2}{(V - V_i)^2 + \gamma_i^2}, \tag{4}$$

Here $c_i$ and $\gamma_i$ are the amplitude and width of the peaks. The values of the partial widths, $\Gamma^{L/R}$, in Eq. (3) are extracted from the fit of the data to Eq. (4). The small amplitudes of the defect peaks in Fig. 4a (corresponding to $G \ll \frac{2e}{h}$ in Eq. (1)) imply an asymmetric location of the defect, such that either $\Gamma_i^L \ll \Gamma_i^R$ or $\Gamma_i^R \ll \Gamma_i^L$. To distinguish between the two possibilities, we can use the fact that, if $d_L = d_R$, one should observe identical peaks, symmetric around $V = 0$, at positive and negative bias $V_i = (\epsilon_i - \epsilon_F)\frac{d_L+d_R}{ed_L}$ and $\tilde{V}_i = -(\epsilon_i - \epsilon_F)\frac{d_L+d_R}{ed_R}$ respectively. As no peak is observed in the range -0.12V $< V <$ 0 in experiment, we conclude that $d_L > d_R$ and $\Gamma_i^L \ll \Gamma_i^R$. The values of $\Gamma^{L/R}$ are then evaluated as

$$\Gamma_i^R = \frac{ed_L \gamma_i}{d_R + d_L}, \tag{5}$$

$$\Gamma_i^L = \Gamma_i^R \frac{hc_i(d_L + d_R)}{4e^2 d_L N_i}, \tag{6}$$

where $N_i$ is the number of equivalent defects of type $i$. Without loss of generality, here we consider $N_i = 1$. The fit to the data yields two Lorentzian peaks in the vicinity of $V_b = 0.12$ V, located at 0.124 ± 0.02 V, 0.138 ± 0.02 V for parallel magnetization (widths $\gamma_1 = 0.0084$ V, $\gamma_2 = 0.0080$ V and amplitudes $c_1 = 0.11$ μS, $c_2 = 0.064$ μS, respectively), and at 0.130 ± 0.02 V, 0.140 ± 0.02 V for anti-parallel magnetization (widths $\gamma_1 = 0.0082$ V, $\gamma_2 = 0.0065$ V and amplitudes $c_1 = 0.093$ μS, $c_2 = 0.067$ μS, respectively), see Fig.4b. The different amplitudes of the two peaks are naturally explained by being related to different spin states, since the density of states in ferromagnetic electrodes is spin dependent. For the same reason, the difference in amplitudes should decrease in the anti-parallel configuration, as indeed observed.

To get a quantitative estimate for the location of the defect, we use Simmons' model [S3] to fit the measured differential conductance at moderate bias 0.1 V $< |V| <$ 0.4 V. Assuming the effective barrier thickness to be equal to 1.3 nm (thickness of trilayer hBN plus separation to the FM electrodes) leads to effective electron mass $m = 0.5\, m_e$, barrier height $\varphi_0 = 3.8$ eV and tunneling exponent $2\kappa = 13$ nm$^{-1}$, corresponding to the conductance decreasing by a factor of $e^{-2\kappa c}$ ~ 1/50 per hBN layer (with $c = 0.3$ nm the distance between hBN layers), in agreement with the results of ref. [S4,S5]. Within this tunneling model, the location of the defect is related to the asymmetry between $\Gamma^L$ and $\Gamma^R$ as

$$\frac{\Gamma^R}{\Gamma^L} = e^{2\kappa(d_L - d_R)}, \qquad (7)$$

if we assume that the two ferromagnetic electrodes are identical. Solving Eqs. (5,7) self-consistently leads to

$$\frac{d_R}{d_L} \approx \frac{3}{7}, \qquad d_L - d_R \approx 0.46 \text{ nm}, \qquad \Gamma^R \approx 2.6 \cdot 10^{-3} \text{ eV}, \qquad \Gamma^L \approx 5.2 \cdot 10^{-6} \text{ eV}. \qquad (8)$$

This supports our assumption $\Gamma^L \ll \Gamma^R$ and shows that the defect is not located at the center of the barrier but is probably located between two of the three hBN layers (dissimilar FM electrodes can also contribute to the asymmetry between $\Gamma_i^R$ and $\Gamma_i^L$ but to a lesser degree). One candidate for such an interlayer defect is a boron monovacancy, which are known to be often present in hBN and form stable configurations via either one or two interlayer covalent bonds [S6].

As seen in Fig. 4a (main text), the splitting of defect levels is larger for parallel magnetization of the electrodes, as the spin-related effects of both electrodes add up. This clearly indicates that the defect is magnetic. As suggested in ref. [S2], the splitting of defect energy levels may be caused by exchange or super-exchange interaction of a quasi-bound defect state with electrons in the metallic electrodes, becoming significant for defect levels in narrow insulator junctions. The above estimate for the location of the defect allows us to quantify the magnitude of exchange splitting induced by the ferromagnetic electrodes in our devices. Using the fitted bias voltage corresponding to the peaks, $V_{i,\mu}$, we find the induced spin splitting at the defect location, $\Delta = \epsilon_{i,\downarrow} - \epsilon_{i,\uparrow}$, for the parallel and anti-parallel configurations to be,

$$\Delta^P \approx 9.8 \text{ meV}, \qquad \Delta^{AP} \approx 7.0 \text{ meV}.$$

Assuming that the above spin splitting is the result of linear superposition of the effect from the two FM electrodes (add up for parallel and subtract for antiparallel magnetization) we obtain that the right electrode contributes $\Delta^R$~ 8.4 meV while the left electrode contributes $\Delta^L$~ 1.4 meV. We can then use these values and the ratio of the defect separations from the two FMs, $d_R/d_L \approx 3/7$, to estimate the decay length for the proximity-induced spin splitting. This yields a slower decay rate with distance (~ 1/5 per hBN layer) compared to the tunneling conductance (~ 1/50 per hBN layer, see above). We note that such a slow decay of proximity-induced spin splitting by hBN/(Co,Ni) interfaces has been predicted theoretically both for the case of Dirac states in graphene/hBN stacks and for the conduction/valence bands of the hBN itself [S7]. In particular, for the case of proximity for individual hBN layers, a decay rate of ~ 1/10 per hBN layer and spin splitting in the second layer of 10—100 meV were predicted [S7], with the lower limit in good agreement with up to 10 meV spin splitting found in this work.

The different spin splitting and bias dependences for parallel and antiparallel magnetization, as seen in the bottom inset of Fig. 4a, lead to a sharp peak in MR at 0.12 V followed by a dip at higher $V$, where the conductance peak for the anti-parallel magnetization is located. In other words, the modulation of the MR occurs due to the difference in proximity-induced spin splitting, $\Delta^P - \Delta^{AP} \sim 3$ meV. At elevated temperatures, $T \gg 10$ K, inelastic processes and thermal broadening start to dominate and the MP peak disappears, as shown in the top panel inset of Fig. 4a.

2. **Temperature dependence of hBN-based MTJs**

To elucidate the behavior of our MTJs, we measured the temperature dependence of the junction resistance, which was not reported for tunnel junctions with an hBN barrier in earlier studies. As shown in Fig. 1b in the main text, for our 3-layer hBN device, $R(T)$ is metallic-like, decreasing by ~20% from room temperature to ~50 K and becoming almost temperature independent between 50 and 10 K. We have carried out this characterization for all our hBN-based devices, including the devices discussed in the main text (see Fig. S2a), plus two other devices discussed in Section 4 of this Supplementary Information (see Fig. S2b).

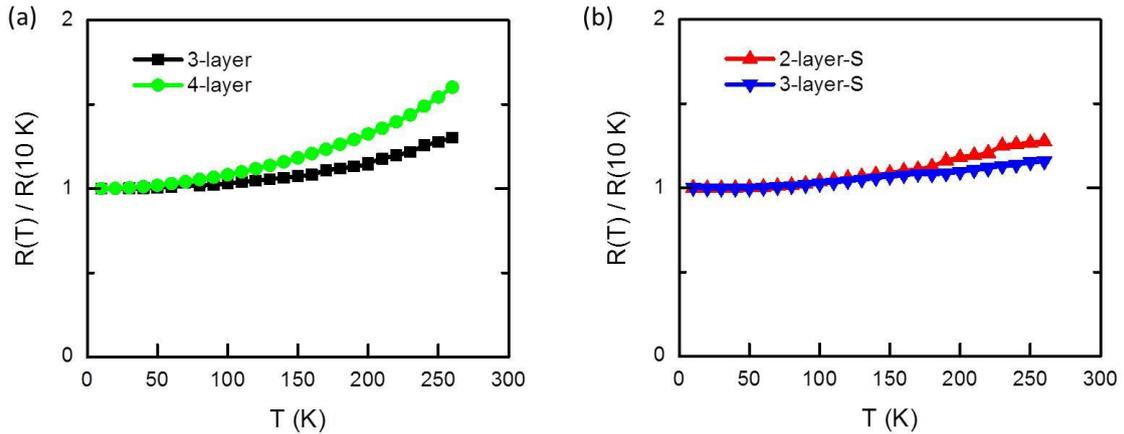

FIG. S2. **Dependence of junction resistance, $R(T)$**. (a) Resistance for two samples with data shown and discussed in the main text, a 3-layer hBN sample, with $R(10$ K$) = 215$ kΩ, and a 4-layer hBN sample, with $R(10$ K$) = 1.2$ MΩ. (b) Resistance for two other samples for which data is shown in Section 4 of this Supplementary Information, a 2-layer hBN sample, with $R(10$ K$) = 4$ kΩ, and a 3-layer hBN sample, with $R(10$ K$) = 252$ kΩ. For all samples $R(T)$ drops as $T$ decreases.

All our devices exhibit a similar behavior: an almost constant or even metallic-like temperature dependence of the junction resistance, as shown in Fig. S2. This behavior, together with the observation of hBN phonon peaks by inelastic electron tunneling spectroscopy, evidence that transport in our devices is not dominated by polymer residues (which lead to insulating-like temperature dependence) but by tunneling via the hBN barrier.

### 3. Inelastic electron tunneling spectroscopy (IETS)

We analyzed the *I-V* characteristic in the full bias range by studying its second derivative, $d^2I/dV^2$, for our 3-layer hBN device. As shown in Fig. 3 in the main text, we extracted its antisymmetric response and performed a multi-peak Gaussian fit, which allows a quantitative comparison with hBN-based tunneling devices from literature. In Table S1, we directly compare the positions of our fitted IETS peaks to those of a graphene-hBN-graphene heterostructure with dominant hBN features (Device 2 in ref. [S8]). All fitted peaks are located within ~10 mV of the bias values reported in ref. [39], with an excellent agreement within ~3 mV for the four main peaks labelled (v) to (viii) located at $V_b < 0.10$ V. These features are consistent with van Hove-like peaks in the single phonon density of states of hBN at energies between ~10 and 190 meV, as found both theoretically and experimentally [S8,S9]. The fact that the IETS spectrum is consistent with hBN phonon peaks is by itself a strong proof that transport in our junctions is not dominated by metallic pinholes nor by polymer residues, but by (phonon-assisted) tunneling via the hBN barrier.

| Peak no. | IETS ref. [S8] | 3-layer fit | Difference |
|---|---|---|---|
| i | 189 | 182 | -8 |
| ii | 163 | 160 | -3 |
| iii | 133 | 121 | -12 |
| iv | 109 | 96 | -14 |
| v | 80 | 77 | -4 |
| vi | 59 | 55 | -4 |
| vii | 36 | 37 | 1 |
| viii | 12 | 15 | 3 |

TABLE S1. Comparison of IETS peak positions (in meV) of 3-layer hBN device with Device 2 from ref. [S8].

As mentioned in the main text, the IETS spectrum from the reference 4-layer hBN device yielded a similar result to that of the 3-layer device. To make a direct comparison between both results, we present in Fig. S3 both the $dV/dI$ and $d^2I/dV^2$ spectroscopies for the reference 4-layer device, together with the multi-peak Gaussian fit from the 3-layer device. Direct agreement between all dominant features in both IETS spectra in Fig S3b confirms the consistent observation of phonon-assisted tunneling via hBN for both devices discussed in the main text.

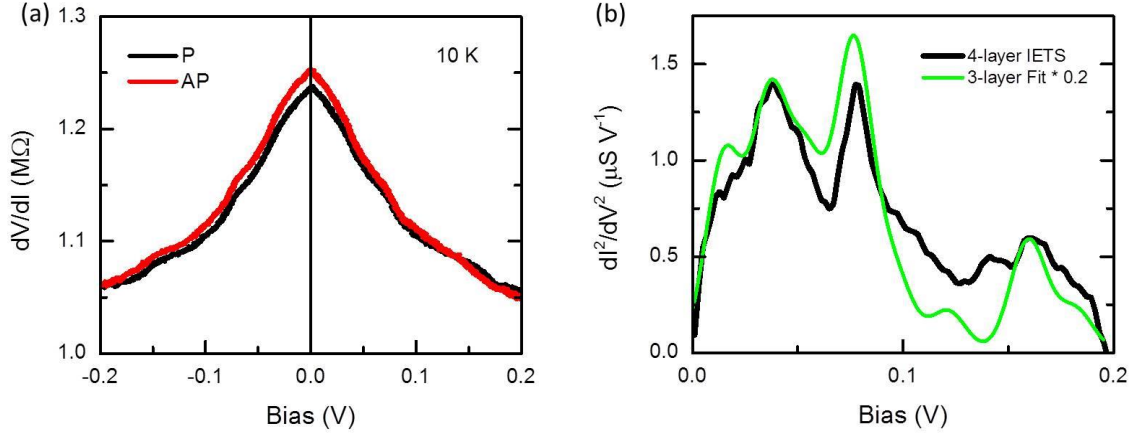

FIG. S3. **Bias spectroscopy and analysis via inelastic electron tunneling spectroscopy (IETS) of 4-layer hBN sample**. (a) Differential resistance (d$V$/d$I$) of the MTJ for the parallel (P) and antiparallel (AP) configuration of the ferromagnetic electrodes. Several kinks can be seen, see also inset of Fig. 2a in the main text. (b) Antisymmetric IETS component. The IETS response for both magnetic configurations is similar so they have been averaged (black line). For comparison, we also show the multi-peak fit to the IETS of the 3-layer hBN sample (green line) previously shown in the main text (top panel of Fig. 3b).

### 4. Tunneling spectroscopy and magnetoresistance data for two other devices

To ensure reproducibility of the data on different devices (in addition to those discussed in the main text), we have fabricated and studied two more devices: a 2-layer device, that showed relatively clean characteristics, similar to the 4-layer device presented in the main text; and another 3-layer device, showing bias-dependent features very similar to the 3-layer device in Figs. 1c and 2. The data for the 2-layer hBN device is shown in Fig. S4. It exhibits a differential resistance, d$V$/d$I$, with several smooth kinks, similar to the spectra for the devices discussed in the main text. Importantly, there are no strong features either in d$V$/d$I$ or in the differential magnetoresistance. This is similar to the reference 4-layer device in the main text.

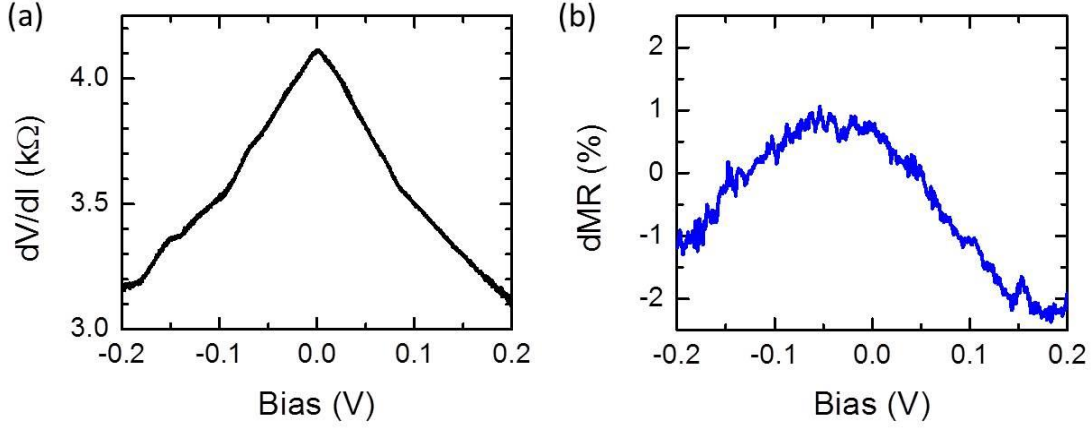

FIG. S4. **Resistance and magnetoresistance of a 2-layer hBN sample**. (a) Differential resistance d$V$/d$I$ in the parallel (P) configuration. (b) Differential MR for the same device. No resonant feature can be seen. All data for $T$ = 10 K.

In contrast, the second 3-layer hBN device shows strong features present at finite bias for both polarities: one at +0.07 V and another -0.09 V (see Fig. S5), with very similar phenomenology to the 3-layer device in the main text. For charge transport, the signature of defect-assisted tunneling is a clear dip in differential resistance, d$V$/d$I$, for both bias values, +0.07 V and -0.09 V. Importantly, we also see the corresponding features in magnetoresistance. As shown in Fig S5b, at low bias both the DC magnetoresistance and the differential magnetoresistance are ~0.5% while at +0.07 V and -0.09 V, the DC magnetoresistance (MR) increases to ~1% and the differential magnetoresistance (dMR) shows prominent features with values of ca. ±1% relative to the background.

The enhancement in MR and dMR in this second 3-layer device is larger, both in absolute and relative terms, compared to the 3-layer hBN device presented in the main text. For example, in absolute terms, the device in the main text shows an increase in MR of ~0.1%, whereas the device presented in Fig. S5 shows an increase of ~0.5%. To understand this relative enhancement, we have analyzed the data of Fig. S5 using the model presented in Section 1.

Given that both (d)MR features at +0.07 V and -0.09 V are similar, yielding an almost symmetric response around $V = 0$ bias, we associate both features with a single defect state. Within this framework and following the model in Section 1, we identify the following relations: $V_i = +0.07\,V = (\epsilon_i - \epsilon_F)\frac{d_L+d_R}{ed_L}$ and $\tilde{V}_i = -0.09\,V = -(\epsilon_i - \epsilon_F)\frac{d_L+d_R}{ed_R}$, respectively. Similar to the previous 3-layer hBN sample, we again conclude that $d_L > d_R$ and $\Gamma_i^L \ll \Gamma_i^R$. Nevertheless, now we can directly extract both the spatial location of the defect within the barrier and its energy alignment relative to the Fermi level. The results, $d_L - d_R \approx 0.16$ nm and $\epsilon_i - \epsilon_F = 39$ meV, are consistent with a defect which is both closer to the center of the barrier and to the Fermi level, as compared to the 3-layer device from the main text (with values of 0.46 nm and 75 meV, respectively). This indicates that MR enhancement via impurity-assisted resonant tunneling can be amplified when such defect is closer to the center of the barrier and to the Fermi energy, opening a pathway to engineer such defect states.

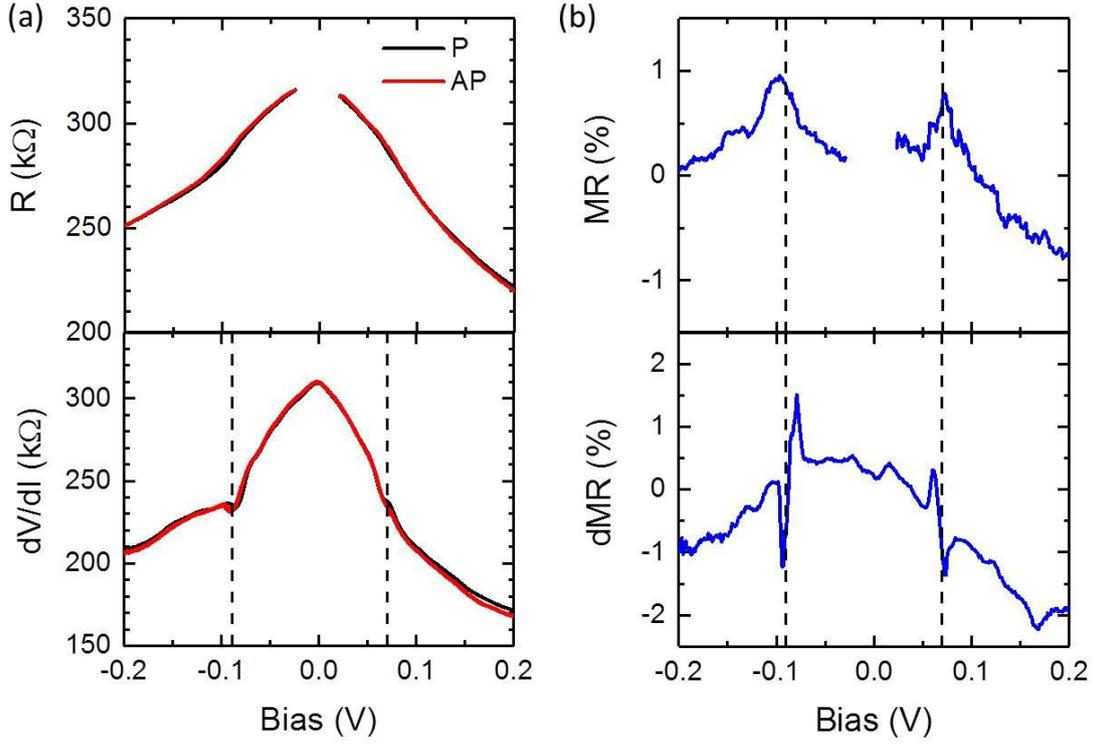

FIG. S5. **Resistance and magnetoresistance for another 3-layer hBN sample**. (a) Bias spectroscopy. Top panel: $R(V)$ for the parallel (P) and antiparallel (AP) configuration of the ferromagnetic electrodes. Bottom panel: Differential resistance $dV/dI(V)$ of the MTJ. Two distinct features can be seen at opposite bias polarity (marked by the vertical dotted lines). (b) Top panel: DC magnetoresistance MR = $(R_{AP}-R_P)/R_P$ with strong maxima of ~1 % observed at finite bias for both polarities, marked by the vertical dotted lines. Bottom panel: differential MR for the same device, derived from $dV/dI_P$, $dV/dI_{AP}$ in (a). Strong changes of differential MR occur at both bias polarities, marked by the vertical dotted lines. All data at $T$ = 10 K.